\documentclass{elsart}
\usepackage{epsfig}

\begin{document}

\begin{frontmatter}

\title{Exponential and power-law probability distributions of wealth
    and income in the United Kingdom and the United States}

\author{Adrian Dr\u{a}gulescu, Victor M. Yakovenko\thanksref{email}}

\thanks[email]{yakovenk@physics.umd.edu,   
   http://www2.physics.umd.edu/\~{}yakovenk}

\address{Department of Physics, University of Maryland, College Park,
  MD 20742-4111, {\bf cond-mat/0103544, v.2 March 28, 2001} \hfill USA}

\begin{abstract}
  We present the data on wealth and income distributions in the United
  Kingdom, as well as on the income distributions in the individual
  states of the USA.  In all of these data, we find that the great
  majority of population is described by an exponential distribution,
  whereas the high-end tail follows a power law.  The distributions
  are characterized by a dimensional scale analogous to temperature.
  The values of temperature are determined for the UK and the USA, as
  well as for the individual states of the USA.
\end{abstract}


\end{frontmatter}

\section{Introduction}

The study of wealth and income distributions has a long history.
Pareto \cite{Pareto} proposed in 1897 that wealth and income
distributions obey universal power laws.  Subsequent studies have
often disputed this conjecture (see a systematic survey in the World
Bank research publication \cite{Kakwani}).  Mandelbrot
\cite{Mandelbrot} proposed that the Pareto law applies only
asymptotically to the high ends of the distributions.  Many
researchers tried to deduce the Pareto law from a theory of stochastic
processes.  Gibrat \cite{Gibrat} proposed in 1931 that income and
wealth are governed by multiplicative random processes, which result
in a log-normal distribution.  These ideas were later followed, among
many others, by Montroll and Shlesinger \cite{Montroll}.  However,
already in 1945 Kalecki \cite{Kalecki} pointed out that the log-normal
distribution is not stationary, because its width increases in time.
Modern econophysicists \cite{Solomon,Bouchaud,Sornette} also use
various versions of multiplicative random processes in theoretical
modeling of wealth and income distributions.

Unfortunately, numerous recent papers on this subject do very little
or no comparison at all with real statistical data, much of which is
widely available these days on the Internet.  In order to fill this
gap, we analyzed the data on income distribution in the United States
(US) from the Bureau of Census and the Internal Revenue Service (IRS)
in Ref.\ \cite{DY-income}.  We found that the individual income of
about 95\% of population is described by the exponential law.  The
exponential law, also known in physics as the Boltzmann-Gibbs
distribution, is characteristic for a conserved variable, such as
energy.  In Ref.\ \cite{DY-money}, we argued that, because money
(cash) is conserved, the probability distribution of money should be
exponential.  Wealth can increase or decrease by itself, but money can
only be transferred from one agent to another.  So, wealth is not
conserved, whereas money is.  The difference is the same as the
difference between unrealized and realized capital gains in stock
market.

Unfortunately, we were not able to find data on the distribution of
money.  On the other hand, we found data on wealth distribution in the
United Kingdom (UK), which are presented in this paper.  Also
presented are the income distribution data for the UK and for the
individual states of the USA.  In all of these data, we find that the
great majority of population is described by an exponential
distribution, whereas the high-end tail follows a power law.

\section{Wealth distribution in the United Kingdom}

In this section, we discuss the cumulative probability distribution of
wealth $N(w)$=(the number of people whose individual wealth is greater
than $w$)/(the total number of people).  A plot of $N$ vs.\ $w$ is
equivalent to a plot of person's rank in wealth vs. wealth, which is
often used for top reach people \cite{Solomon97}.  We will use the
power law, $N(w)\propto 1/w^\alpha$, and the exponential law
$N(w)\propto\exp(-w/W)$, to fit the data.  These distributions are
characterized by the exponent $\alpha$ and the ``temperature'' $W$.
The corresponding probability densities, $P(w)=-dN(w)/dw$, also follow
a power law or an exponential law.  For the exponential law, it is
also useful to define the temperatures $W^{(2)}$ (also known as the
median) and $W^{(10)}$ using the bases of 1/2 and 1/10:
$N(w)\propto(1/2)^{w/W^{(2)}}\propto(1/10)^{w/W^{(10)}}$.

\begin{figure}
\centerline{
  \epsfig{file=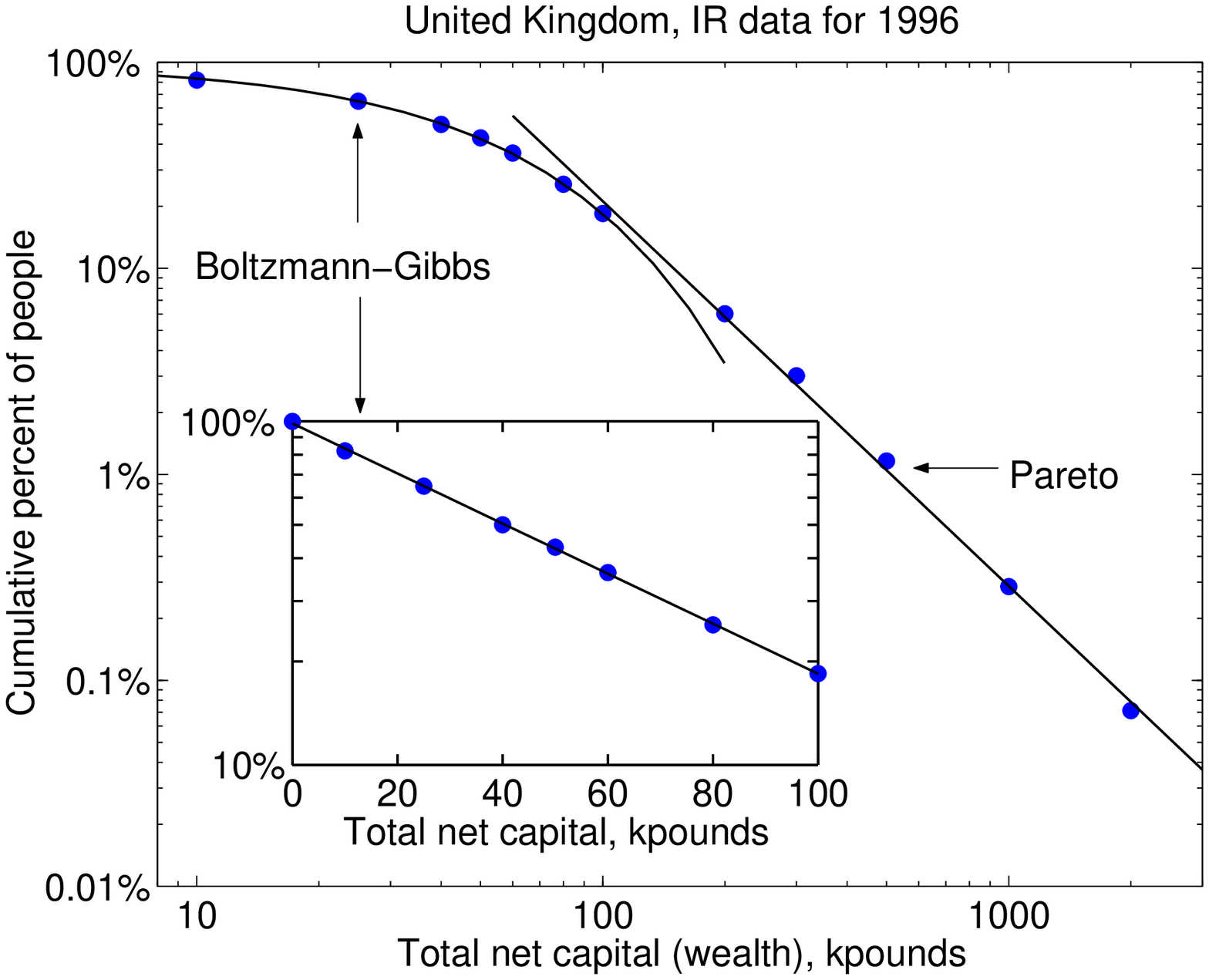,width=0.54\linewidth}
  \epsfig{file=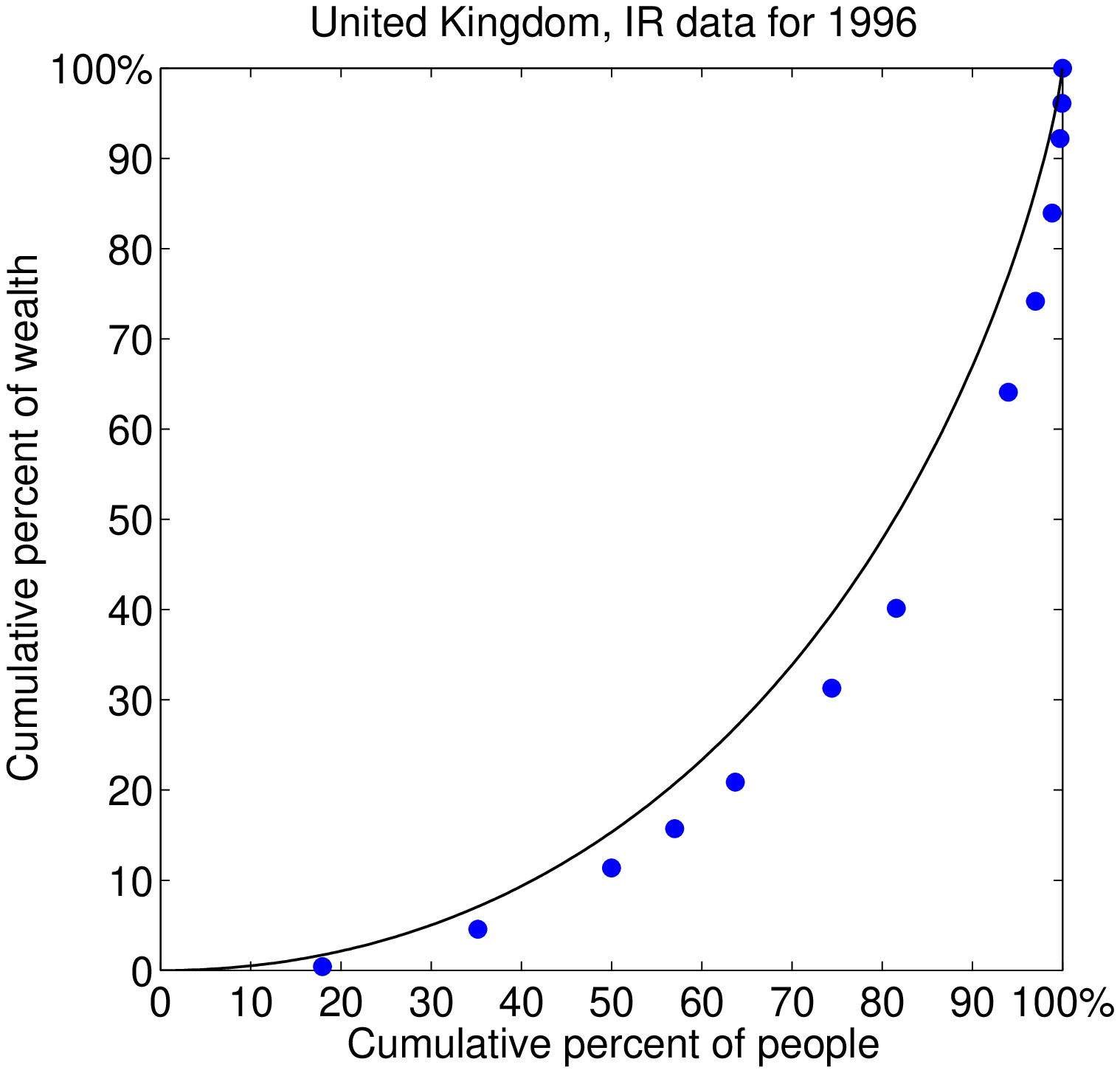,width=0.455\linewidth}}
\caption{\footnotesize\sf Cumulative probability distribution of total
  net capital (wealth) shown in log-log (left panel), log-linear
  (inset), and Lorenz (right panel) coordinates.  Points: the actual
  data.  Solid lines: left panel -- fits to the exponential
  (Boltzmann-Gibbs) and power (Pareto) laws, right panel -- function
  (\ref{eq:yx}) calculated for the exponential law.}
\label{fig:UKwealth}
\end{figure}

The distribution of wealth is not easy to measure, because people do
not report their total wealth routinely.  However, when a person dies,
all assets must be reported for the purpose of inheritance tax.  Using
these data and an adjustment procedure, the British tax agency, the
Inland Revenue (IR), reconstructed wealth distribution of the whole UK
population.  In Fig.\ \ref{fig:UKwealth}, we present the 1996 data
obtained from their Web site \cite{UKwealth}.  The left panel shows
the cumulative probability as a function of the personal total net
capital (wealth), which is composed of assets (cash, stocks, property,
household goods, etc.)  and liabilities (mortgages and other debts).
The main panel illustrates in the log-log scale that above
100~k$\pounds$ the data follow a power law with the exponent
$\alpha=1.9$.  The inset shows in the log-linear scale that below
100~k$\pounds$ the data is very well fitted by an exponential
distribution with the temperature $W_{UK}=59.6$~k$\pounds$
($W_{UK}^{(2)}=41.3$~k$\pounds$ and $W_{UK}^{(10)}=137.2$~k$\pounds$).

The right panel of Fig.\ \ref{fig:UKwealth} shows the so-called Lorenz
curve \cite{Kakwani,DY-income} for wealth distribution.  The horizontal and
vertical coordinates are the cumulative population $x(w)$ and the
cumulative wealth $y(w)$:
\begin{equation}
x(w)=\int_0^wP(w')\,dw', \qquad
y(w)=\int_0^w w'P(w')\,dw'\,/\,\int_0^\infty w'P(w')\,dw'.
\end{equation}
The points represent the actual data, whereas the solid line is
calculated for a purely exponential distribution \cite{DY-income}:
\begin{equation}
    y = x +(1-x)\ln(1-x).
\label{eq:yx}
\end{equation}
One can see that the data systematically deviate from the exponential
law because of the wealth concentrated in the power-law tail; however,
the deviation is not very big.  The so-called Gini coefficient
\cite{Kakwani,DY-income}, which measures the inequality of wealth
distribution, has increased from 64\% in 1984 to 68\% in 1996
\cite{UKwealth}.  This value is bigger than the Gini coefficient 50\%
for a purely exponential distribution \cite{DY-income}.  The
inequality of the US income distribution was also increasing during
that time period \cite{DY-income}.

\section{Income distribution in the United Kingdom}

\begin{figure}
\centerline{
  \epsfig{file=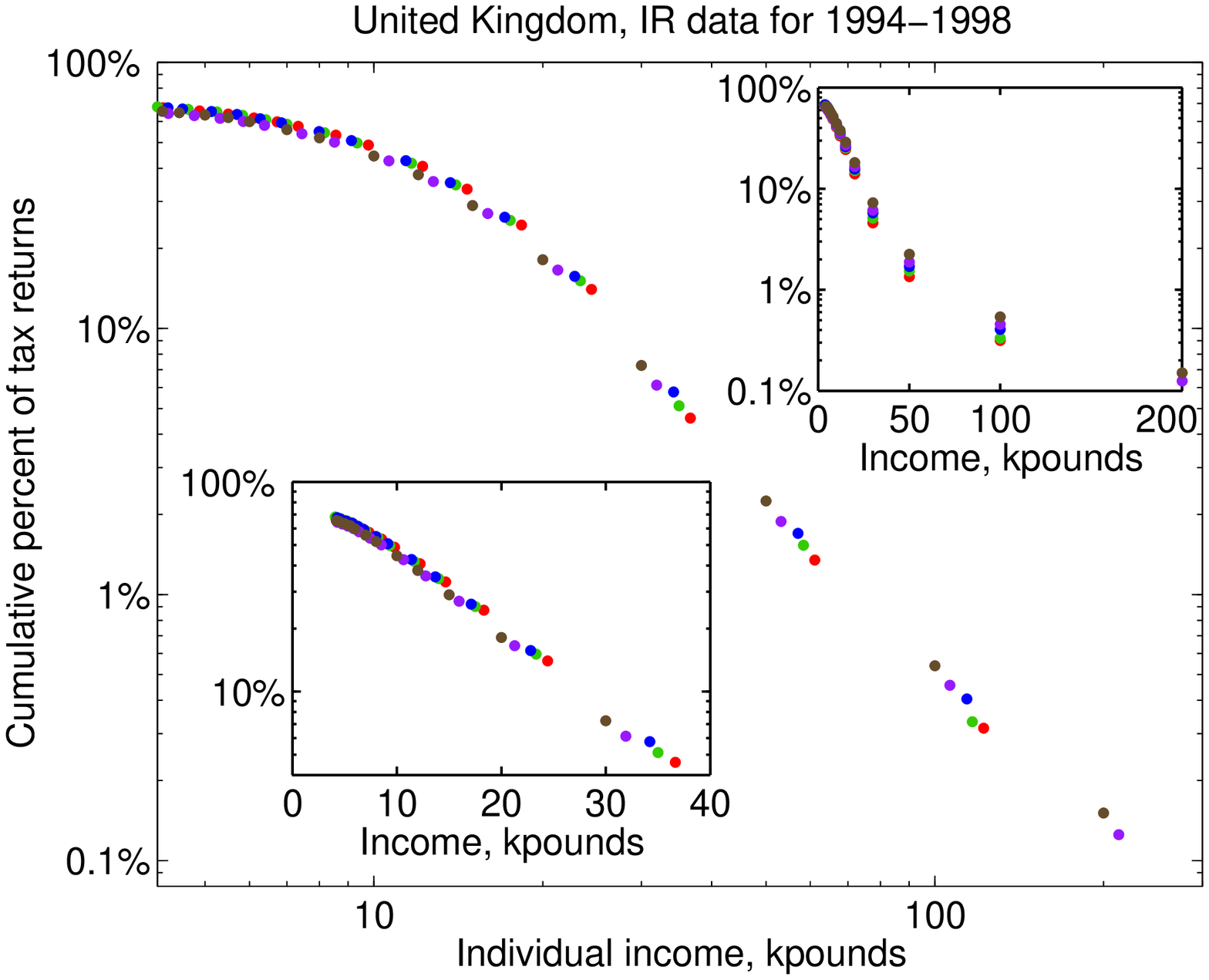,width=0.545\linewidth}
  \epsfig{file=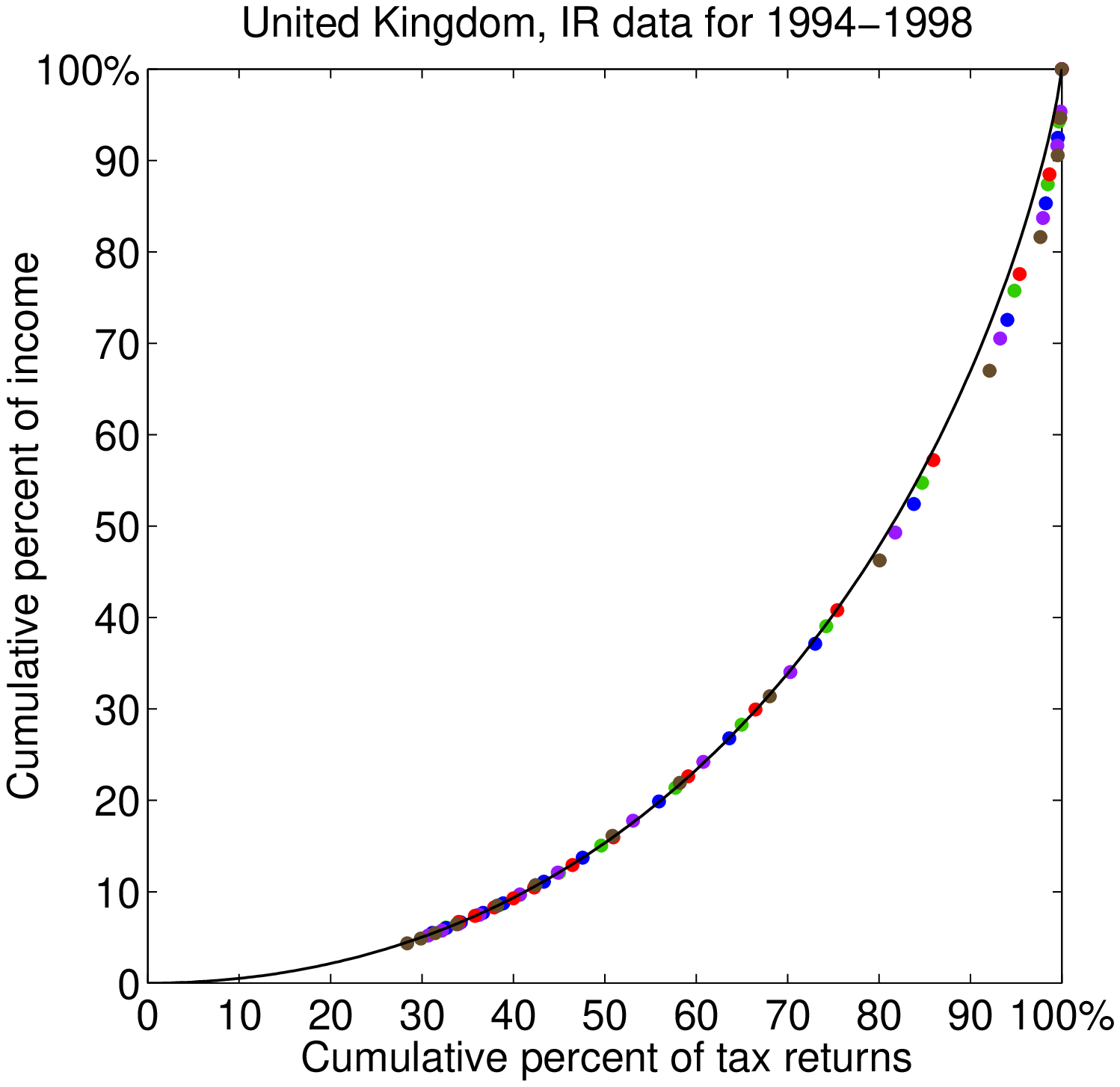,width=0.445\linewidth}}
\caption{\footnotesize\sf Cumulative probability distributions of
 yearly individual income in the UK shown as raw data (top inset) and
 scaled data in log-log (left panel), log-linear (lower inset), and
 Lorenz (right panel) coordinates.  Solid curve: fit to function
 (\ref{eq:yx}) calculated for the exponential law.}
\label{fig:UKincome}
\end{figure}

We obtained the data on the yearly income distribution in the UK for
1997/8 and 1998/9 from the Web site of the IR \cite{UKincome}.  The
data for 1994/5, 1995/6, and 1996/7 were taken from the Annual
Abstract of Statistics derived from the IR \cite{ONS}.  The data for
these 5 years are presented graphically in Fig.\ \ref{fig:UKincome}.
In the upper inset of the left panel, the original raw data for the
cumulative distribution are plotted in log-linear scale.  For not too
high incomes, the points form straight lines, which implies the
exponential distribution $N(r)\propto\exp(-r/R)$, where $r$ stands for
income (revenue), and $R$ is the income ``temperature''.  However, the
slopes of these lines are different for different years.  The
temperatures for the years 1994/5, 1995/6, 1996/7, and 1997/8 differ
from the temperature for 1998/9, $R_{UK}^{(98/9)}=11.7$~k$\pounds$
($R_{UK}^{(2)}=8.1$~k$\pounds$ and $R_{UK}^{(10)}=26.9$~k$\pounds$),
by the factors 0.903, 0.935, 0.954, and 0.943.  To compensate for this
effect, we rescale the data.  We divide the horizontal coordinates
(incomes) of the data sets for different years by the quoted above
factors and plot the results in log-log scale in the main panel and
log-linear scale in the lower inset.  We observe scaling: the collapse
of points on a single curve.  Thus, the distributions $N_i(r)$ for
different years $i$ are described by a single function $f(r/R_i)$.
The main panel shows that this scaling function $f$ follows a power
law with the exponent $\alpha=2.0$-2.3 at high incomes.  The lower
inset shows that $f$ has an exponential form for about 95\% of
individuals with lower incomes.  These results qualitatively agree
with a similar study by Cranshaw \cite{Cranshaw}.  He proposed that
the $P(r)$ data for lower incomes are better fitted by the Gamma
distribution $\Gamma(r)\propto r^\beta\exp(-r/R)$.  For simplicity, we
chose not to introduce the additional fitting parameter $\beta$.

We must mention that the individuals with income below a certain
threshold are not required to report to the IR.  That is why the data
in the lower inset do not extend to zero income.  We extrapolate the
straight line to zero income and take the intercept with the vertical
axis as 100\% of individuals.  Thus, we imply that the IR data does
not account for about 25-27\% of individuals with income below the
threshold.

The Lorenz curve for the distribution of the UK income is shown in the
right panel of Fig.\ \ref{fig:UKincome}.  We treat the number of
individuals below the income threshold and their total income as
adjustable parameters, which are the horizontal and vertical offsets
of the coordinates origin relative to the lowest known data point.
These parameters are chosen to fit the Lorenz curve for the
exponential law (\ref{eq:yx}) shown as the solid line.  The fit is
very good.  The horizontal offsets are 28-34\%, which is roughly
consistent with the numbers quoted for the lower inset of the left
panel.

\section{Income distribution in the United States}

\begin{figure}
\centerline{
  \epsfig{file=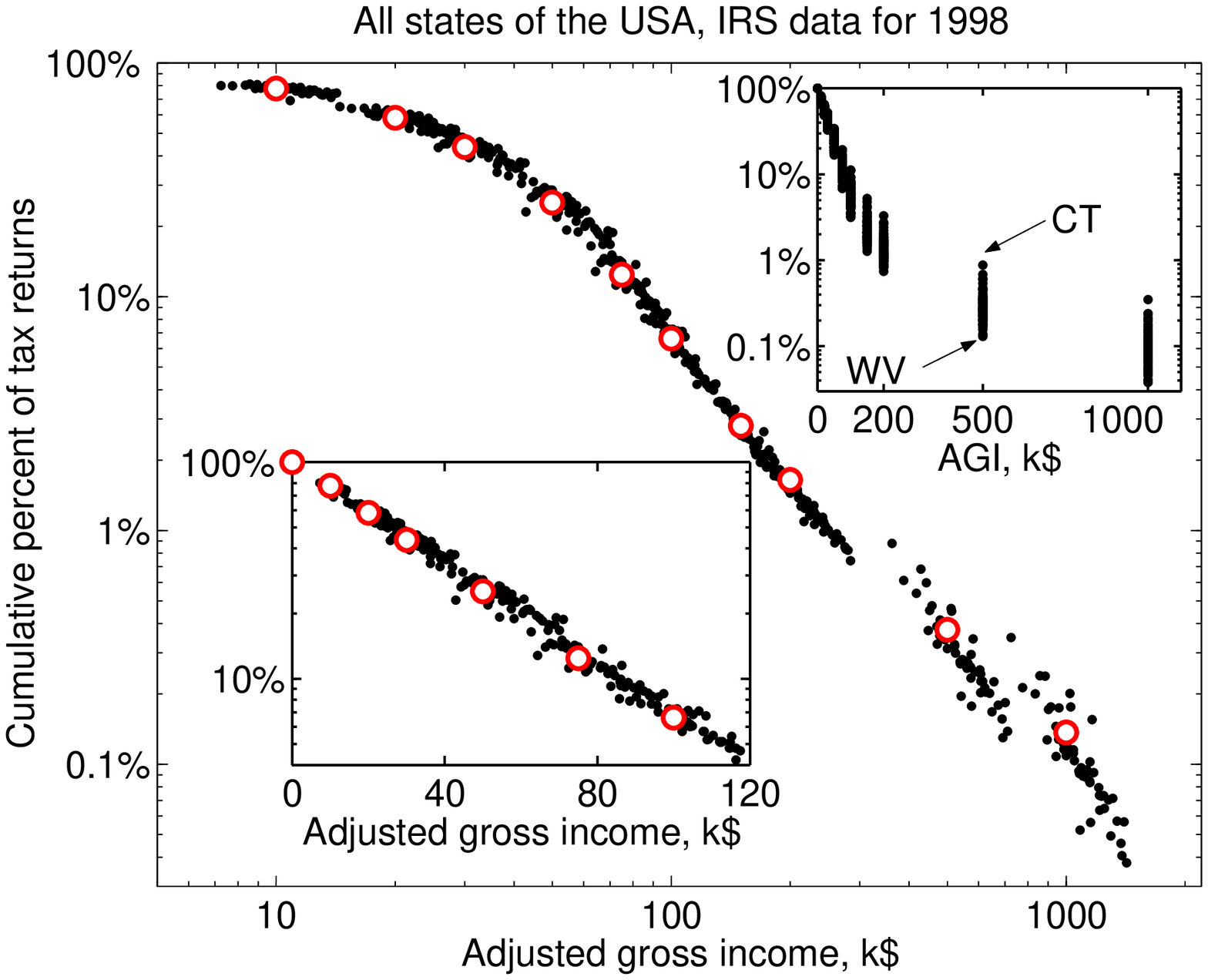,width=0.545\linewidth}
  \epsfig{file=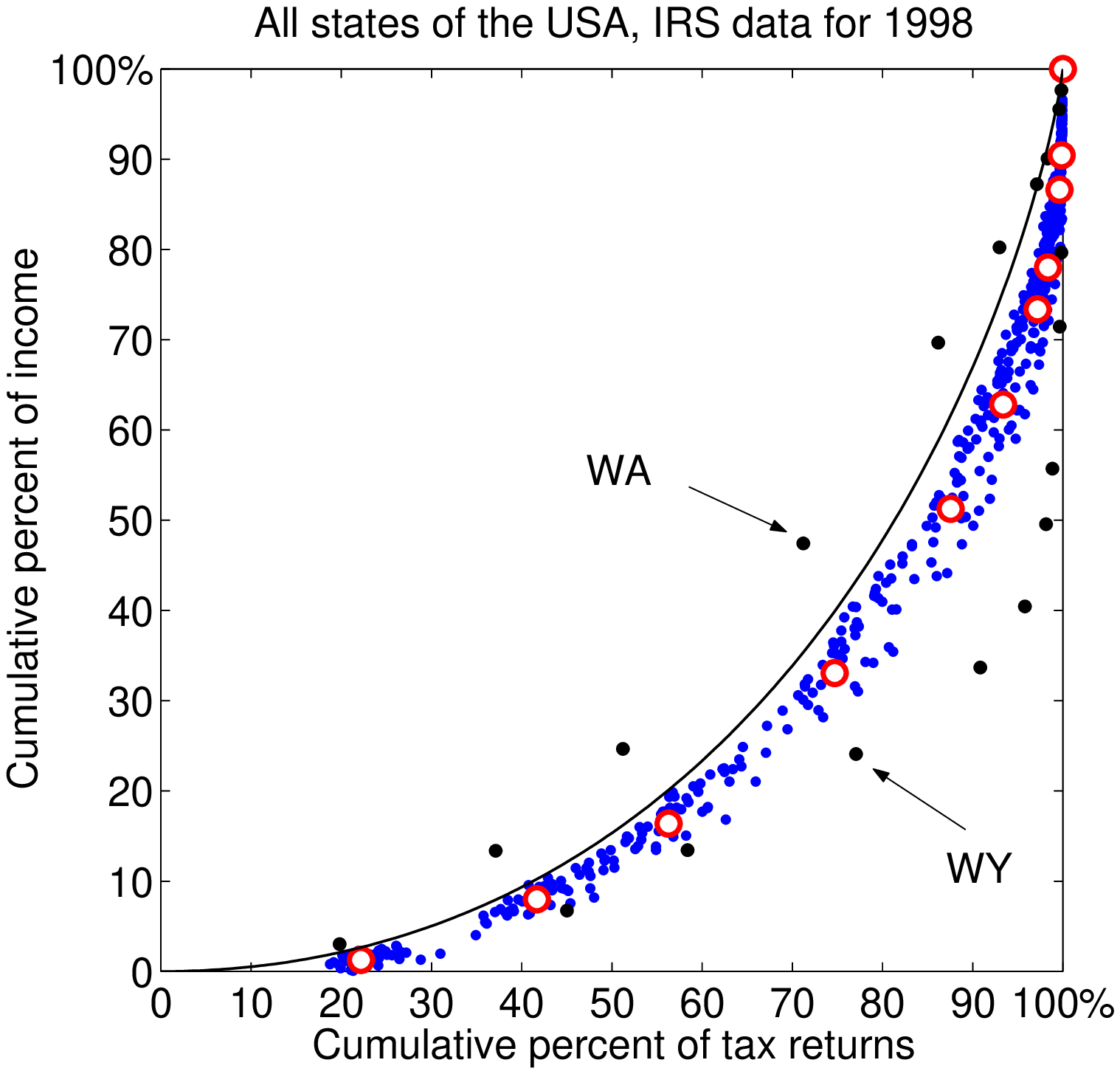,width=0.445\linewidth}}
\caption{\footnotesize\sf Cumulative probability distributions of
 yearly individual income for different states of the USA shown as raw
 data (top inset) and scaled data in log-log (left panel), log-linear
 (lower inset), and Lorenz (right panel) coordinates.  Solid curve:
 function (\ref{eq:yx}) calculated for the exponential law.}
\label{fig:USincome}
\end{figure}

We obtained the data on distribution of the yearly individual income
in 1998 for each of the 50 states and the District of Columbia that
constitute the USA from the Web site of the IRS \cite{IRSsite}.  We
plot the original raw data for the cumulative distribution of income
in log-linear scale in the upper inset of the left panel of Fig.\
\ref{fig:USincome}.  The points spread significantly, particularly at
higher incomes.  For example, the fraction of individuals with income
greater than 1~M\$ varies by an order of magnitude between different
states.  However, after we rescale the data in the manner described in
the preceding section, the points collapse on a single curve shown in
log-log scale in the main panel and log-linear scale in the lower
inset.  The open circles represent the US average, obtained by
treating the combined data for all states as a single set.  We observe
that the distribution of higher incomes approximately follows a power
law with the exponent $\alpha=1.7\pm0.1$, where the $\pm0.1$ variation
includes 70\% of all states.  On the other hand, for about 95\% of
individuals with lower incomes, the distribution follows an
exponential law with the average US temperature $R_{US}=36.4$~k\$
($R_{US}^{(2)}=25.3$~k\$ and $R_{US}^{(10)}=83.9$~k\$).  The
temperatures of the individual states differ from $R_{US}$ by the
amounts shown in Table \ref{table}.  For example, the temperature of
Connecticut (CT) is 1.25 times higher and the temperature of West
Virginia (WV) is 0.78 times lower than the average US temperature.

\begin{table}
\caption{Deviations of the state temperatures from the average US
temperature.}
\scriptsize
\begin{tabular}{@{\extracolsep{-0.8em}}*{17}{c}@{\extracolsep{-0.8em}}}
\hline
 CT  &  NJ  &  MA  &  MD  &  VA  &  CA  &  NY  &  IL  &  CO  &  NH  &  

 AK  &  DC  &  DE  &  MI  &  WA  &  MN  &  GA              \\ [-0.5ex]

25\% & 24\% & 14\% & 14\% &  9\% &  9\% &  7\% & 6\%  &  6\% &  5\% &   

 5\% &  5\% &  4\% &  4\% &  2\% &  1\% &  0\%               \\ \hline
 TX  &  RI  &  AZ  &  PA  &  FL  &  KS  &  OR  &  HI  &  NV  &  NC  &

 WI  &  IN  &  UT  &  MO  &  VT  &  TN  &  NE              \\[-0.5ex]

-1\% & -3\% & -3\% & -3\% & -4\% & -5\% & -6\% & -7\% & -7\% & -7\% &

-8\% & -8\% & -9\% & -9\% & -9\% & -11\% &-12\%             \\ \hline
 OH   &  LA   &  AL   &  SC   &  IA   &  WY   &  NM   &  KY   & ID  & 

 OK   &  ME   &  MT   &  AR   &  SD   &  ND   &  MS   &  WV  \\[-0.5ex]

-12\% & -13\% & -13\% & -13\% & -14\% & -14\% & -14\% & -14\% & -15\% &  

-16\% & -16\% & -19\% & -19\% & -20\% & -20\% & -21\% & -22\% \\ \hline
\end{tabular}
\label{table}
\end{table}

The Lorenz plot for all states is shown in the right panel of Fig.\
\ref{fig:USincome} together with the solid curve representing Eq.\
(\ref{eq:yx}).  The majority of points are well clustered and are not
too far from the solid curve.  The exceptions are Wyoming (WY) with
much higher inequality and the Washington state (WA) with noticeably
lower inequality of income distribution.  The average US data, shown
by open circles, is consistent with our previous results
\cite{DY-income}.  Unlike in the UK case, we did not make any
adjustment in the US case for individuals with income below the
threshold, which appears to be sufficiently low.

\section{Discussion}

We found scaling in the cumulative probability distributions $N(r)$ of
individual income $r$ derived from the tax statistics for different
years in the UK and for different states in the US.  The distributions
$N_i(r)$ have the scaling form $f(r/R_i)$, where the scale $R_i$ (the
temperature) varies from one data set $i$ to another, but the scaling
function $f$ does not.  The function $f$ has an exponential
(Boltzmann-Gibbs) form at the low end, which covers about 95\% of
individuals.  At the high end, it follows a power (Pareto) law with
the exponents about 2.1 for the UK and 1.7 for the US.  Wealth
distribution in the UK also has a qualitatively similar shape with the
exponent about 1.9 and the temperature $W_{UK}=60$~k$\pounds$.  Some
of the other values of the exponents found in literature are 1.5
proposed by Pareto himself ($\alpha=1.5$), 1.36 found by Levy and
Solomon \cite{Solomon97} for the distribution of wealth in the Forbes
400 list, and 2.05 found by Souma \cite{Souma} for the high end of
income distribution in Japan.  The latter study is similar to our work
in the sense that it also uses tax statistics and explores the whole
range of incomes, not just the high end.  Souma \cite{Souma} finds
that the probability density $P(r)$ at lower incomes follows a
log-normal law with a maximum at a nonzero income.  This is in
contrast to our results, which suggest that the maximum of $P(r)$ is
at zero income.  The disrepancy may be due to the high threshold for
tax reporting in Japan, which distorts the data at the low end.  On
the other hand, if the data is indeed valid, it may reflect the actual
difference between the social stuctures of the US/UK and Japan.

The income temperature for the UK in 1998/9 was
$R_{UK}=11.7$~k$\pounds$ and for the US in 1998 was $R_{US}=36.4$~k\$.
Using the exchange rate as of December 31, 1998 to convert pounds into
dollars \cite{currency}, we find that the UK temperature was
$R_{UK}=19.5$~k\$, which is 1.87 times lower than the US temperature.
The difference in temperatures indicates nonequilibrium, which can be
exploited to create a thermal machine \cite{DY-money}.  The gain
(profit) produced by such a thermal machine is proportional to the
difference in temperatures.  In agreement with the second law of
thermodynamics, money would flow from a high-temperature system to a
low-temperature one.  This may explain the huge trade deficit of the
USA in global international trade with other, lower-temperature
countries.  The variation of temperatures between different states of
the USA is shown in Table \ref{table}.

\begin{ack}
We are grateful to Ted Cranshaw, who kindly sent us his unpublished
study of income distribution in the UK \cite{Cranshaw} and data
\cite{ONS}, and Bertrand Roehner, who suggested to study income
distributions in the states of the USA and sent us link
\cite{IRSsite}.
\end{ack}


\end{document}